# Overcoming the indirect bandgap: efficient silicon emission via momentum-expanded photonic states

*Aleksei I. Noskov[1], Alexander B. Kotlyar[2], Liat Katrivas[2], Zakhar Reveguk[2], Evan P. Garcia[1], V. Ara Apkarian[1], Christophe Galland[3], Eric O. Potma[1], Dmitry A. Fishman[1*]*

*[1]Department of Chemistry, University of California, Irvine, Irvine, CA 92697, USA*

*[2]George S. Wise Faculty of Life Sciences, Tel Aviv University, Tel Aviv 6997801, Israel*

*[3]Swiss Federal Institute of Technology, EPFL, Station 3, CH-1015 Lausanne, Switzerland*

## Abstract

Silicon's inherently indirect bandgap severely limits its radiative efficiency, posing a fundamental challenge to the development of practical silicon-based light sources. While strategies such as nanoscale confinement of electrons and holes (quantum dots), Mie resonators, and hybrid plasmonic structures have improved emission, they typically require complex fabrication workflows. Here, we demonstrate a conceptually distinct and scalable approach to enable light emission from a bulk silicon wafer by decorating its surface with gold or copper nanoparticles. Remarkably, the effect is nearly identical for Au and Cu, with particle size emerging as the dominant factor. We show that strong luminescence from the bulk wafer emerges only when the nanoparticle diameter is below 2 nm. We attribute this effect to the formation of spatially confined photonic states with broadened momentum distributions, which must enable diagonal, phonon-independent optical transitions that bypass the limitations imposed by silicon's indirect bandgap. This mechanism yields broadband emission across the visible and near-infrared spectrum, with quantum efficiencies comparable to direct bandgap semiconductors, representing a $10^5$-fold increase in integrated spectral intensity. This discovery challenges the conventional understanding of silicon's optical constraints and opens a practical pathway toward high-performance silicon-based optical and optoelectronic components.

The co-integration of silicon electronics with photonic and optoelectronic components promises transformative advantages, including enhanced speed, increased bandwidth, and reduced power consumption. Silicon, buttressed by its natural abundance and well-established manufacturing infrastructure, holds great promise for photonic integrated circuits [1-3]. However, its indirect bandgap compromises its optical properties. Unlike direct bandgap semiconductors, optical transitions in silicon require phonon assistance to conserve momentum, as illustrated in Figure 1a. This reliance on phonons drastically reduces the rates of optical absorption and emission. Hot carriers rapidly thermalize to states near the band edge within 0.5 ps [4-6]. Due to the slow nature of phonon-assisted radiation, a population bottleneck forms near the bottom of the conduction band, where electron-hole recombination is predominantly governed by non-radiative processes. Consequently, silicon exhibits a low luminescence quantum efficiency ($\eta \sim 10^{-6}$)[7], severely limiting its use as a light emitter.

Several strategies have been proposed to address silicon's low quantum efficiency of light emission. For instance, reducing silicon's dimensions to the nanoscale can increase its luminescence via quantum confinement [8-12]. This effect is generally attributed to an increase in the overlap of electron and hole wavefunctions in momentum space, thereby accelerating radiative direct recombination across the indirect gap. Another approach utilizes (sub)-micrometer-sized silicon resonators, either as isolated structures or arranged in periodic arrays [13-15]. These structures enhance light-matter interaction by leveraging optical resonances, which can increase radiative rates through the Purcell effect or sustain hot electron populations via the Auger effect. Plasmonic enhancements represent an additional avenue for improving silicon's luminescent properties. Integrating silicon with plasmonic nanostructures takes advantage of their strong localized fields and high density of optical states [16-19]. This approach has achieved notable improvements in silicon's light emission efficiency, with reported quantum yields exceeding 1% [16].

Despite recent advancements, practical on-chip silicon light sources remain elusive. This is largely due to fabrication complexities incompatible with existing circuit manufacturing processes. To overcome this, we require approaches that enhance silicon's emission while minimizing material modifications. In this work, we address the fundamental limitation of silicon's indirect bandgap emission: the requirement for phonon-assisted transitions. We demonstrate that by eliminating the need for phonon involvement, silicon's radiative rates can be dramatically enhanced without any alteration, structuring, or modification of the bulk material.

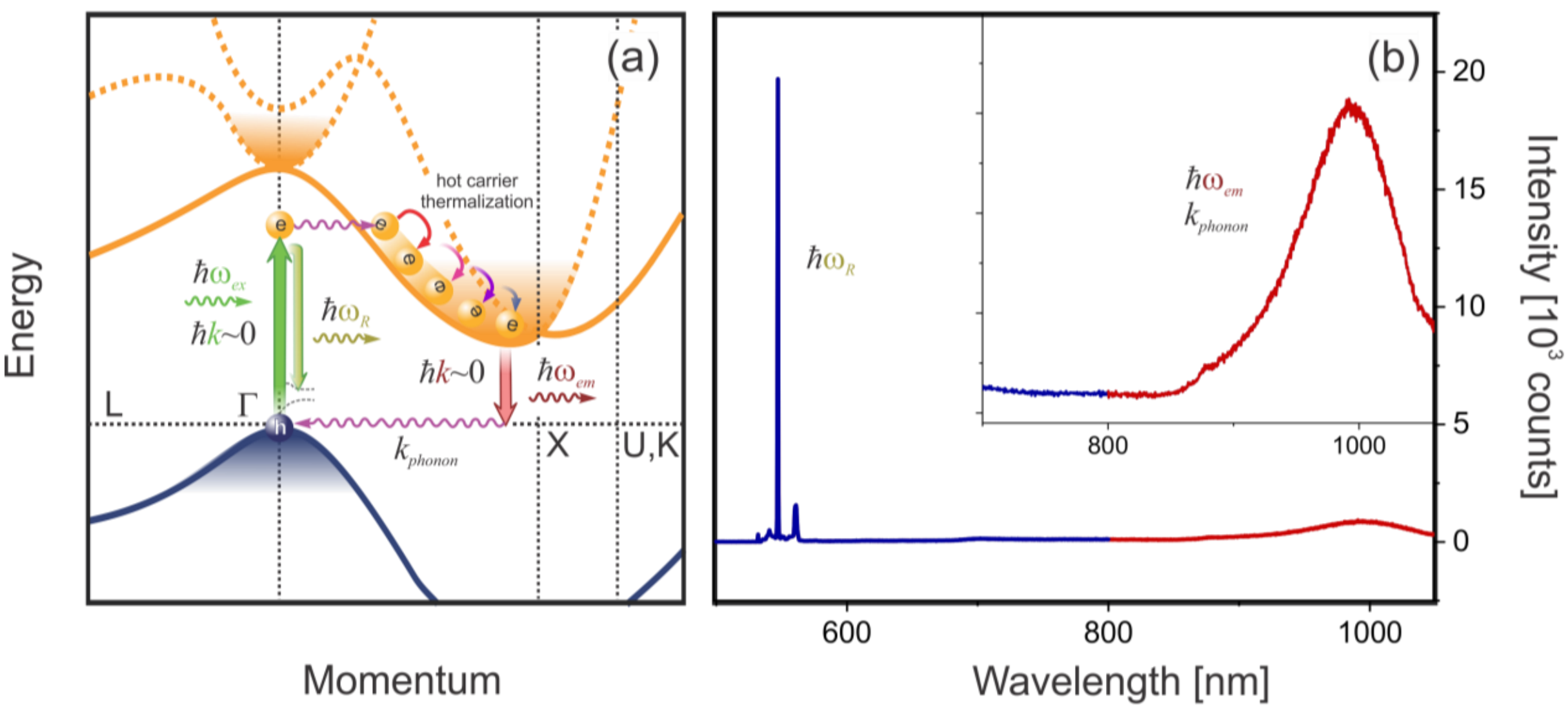

**Figure 1.** (a) Schematic of optical transitions in bulk silicon. Similar to absorption processes, transitions from the bottom of the conduction band require phonon assistance to conserve momentum, leading to the inherently low emission efficiency of bulk silicon. (b) Emission spectrum of bulk silicon, showing contributions from Raman scattering in the visible range (blue) and phonon-assisted emission from the bottom of the conduction band in the near-infrared spectral range (red). Excitation using 532 nm, 0.5 mW, 0.75 NA.

Figure 1b shows the emission spectrum of a clean silicon wafer following laser excitation at 532 nm. In addition to optical phonon Raman lines, the spectrum features weak emission near ~1.0 μm, a phonon-assisted luminescence of rapidly thermalized electrons at the bottom of the conduction band. To overcome silicon's inherently low radiative rates, we apply nanometer-sized gold and copper particles to the silicon's surface. This simple and direct procedure remarkably changes the wafer's emission spectrum in a particle size-dependent manner, as shown in Figure 2a. For 5 nm and 15 nm Au particles, the phonon-assisted luminescence near the silicon band edge remains visible, while an additional spectral band appears in the 600–700 nm range. When the Au nanoparticle size is reduced to 1.2 nm (Supplementary Information Part I), the band-edge luminescence from silicon is fully suppressed and a bright, broadband emission emerges, spanning the entire spectral range from the 532 nm excitation wavelength to approximately 1.0 μm. Nearly identical emission profiles are observed when the silicon surface is decorated with 1.2 nm Cu nanoparticles, indicating that the effect is predominantly dictated by particle size and that the chemical differences between Au and Cu are less of a factor. The observed emission cannot be attributed to intrinsic luminescence from either gold (Figure 2, purple line and discussion in Supplementary Information Part IV) or copper. Instead, it points to the activation of new radiative channels in silicon.

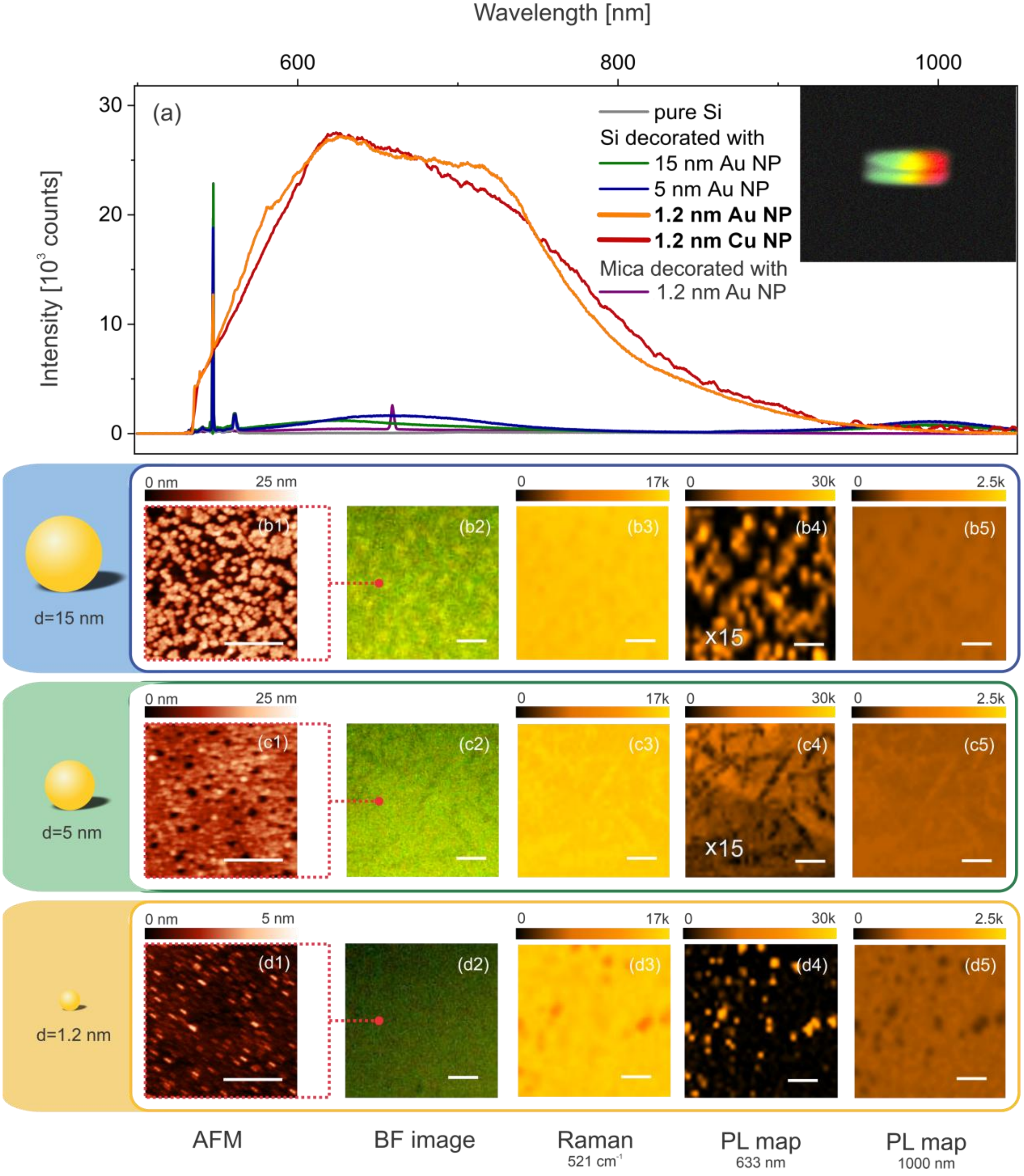

**Figure 2.** (a) Emission spectra of a pure silicon wafer (grey) and wafers decorated with gold and copper nanoparticles: 15 nm Au (blue), 5 nm Au (green), 1.2 nm Au (orange) and 1.2 nm Cu (red, normalized to orange by factor x0.7). Excitation using 532 nm, 0.5 mW, 0.75 NA. The emission spectrum of 1.2 nm Au

particles on mica, representing the intrinsic emission of the nanoparticles themselves and Raman of mica substrate, is shown as the purple spectrum. The inset shows a side view of the emitting silicon wafer captured on a phone camera (10 s exposure) through an Amici prism (see also Figure 4c). Characterization and imaging of wafer surfaces decorated with 15 nm (b), 5 nm (c) and 1.2 nm (d) gold particles: (1) atomic force microscopy maps (scale bar 400 nm), (2) bright-field optical images using white light illumination, (3) Raman maps at the 521 $cm^{-1}$ silicon phonon line, (4) photoluminescence (PL) maps at 630 nm, (5) PL maps at 1000 nm, representing silicon's phonon-assisted luminescence from the bottom of the conduction band at X point. Scale bar is 5 μm for all optical images. Note: PL maps at 630 nm for 15 nm (b4) and 5 nm (c4) samples are enhanced by a factor of 15.

To further examine the nature of the sample's emission characteristics, we performed spatially resolved measurements of the particle-decorated silicon wafers. The atomic force microscopy (AFM) maps in Figures 2b1-d1 confirm the presence of the gold particles on the wafer's surface, forming a monolayer configuration. For the 15 nm particles, the bright-field reflection image in Figure 2b2 reveals faint structures, indicating weak, but visible, plasmon-enhanced scattering activity (see Figure SF3, *Supplementary Information Part II*). Scattering is noticeably weaker for the 5 nm particles, as shown in Figure 2c2, and becomes undetectable for the 1.2 nm Au clusters (Figure 2d2). For these smallest particles, plasmons are overdamped, and scattering is of the Rayleigh type. These experiments underline that, unlike their larger counterparts, the 1.2 nm particles, either Au or Cu (Figure SF4) have no significant effect on the scattering properties of the Si surface and are therefore invisible to the naked eye.

The Raman maps, recorded at the 521 $cm^{-1}$ optical phonon line of silicon, appear largely unaffected by the 5 nm and 15 nm particles deposited on the wafer's surface, as shown in Figures 2b3-c3. For the 1.2 nm decorations, however, the intensity of the silicon's Raman response is reduced, as shown for the patch of particles in Figure 2d3, and consistent with increased optical absorption in the Si surface layer (as discussed further below). A similar trend is seen in the PL maps at silicon's band edge, which are immune to the presence of the larger particles (Figures 2b5-c5), but show a fully depleted signal when 1.2 nm particles are used (Figure 2d5). On the other hand, the location of the Au particles can be clearly seen in the PL map at the 630 nm emission wavelength, with the strongest signals observed for the 1.2 nm particles (Figures 2b4-d4). Similar strong emission is observed for wafer decorated with 1.2 nm Cu particles (Figure 2a, Figure SF7 and SF8, Supplementary Information Part III) Together, the emission maps emphasize that the presence of the sub-2 nm Au or Cu particles induces new emission across the visible and near-IR spectral range, whereas the applied surface decorations do not appear to surface-enhance silicon's Raman lines nor its phonon-assisted PL at the band edge in a meaningful way.

Experiments with 5 nm and 15 nm Au particles on mica exhibit emission features similar to the 600-700 nm band observed in Figure 2a, confirming gold *photoluminescence* (PL)[20-23] as the source of this visible emission (see *Supplementary Information Part IV* for a discussion on the signal's origins from pure Au structures). However, the gold PL from 1.2 nm Au particles on mica is significantly weaker, more than one order of magnitude lower (Figure 2a, purple spectrum, Figure SF11d, *Supplementary Information Part IV*), ruling out gold luminescence as the primary mechanism responsible for the strong broadband emission observed with 1.2 nm Au particles on silicon. Furthermore, the intensity of visible luminescence from silicon wafers decorated with 1.2 nm Au particles remains unchanged regardless of the number of applied particle layers. Even with ~40 layers, the luminescence intensity remains comparable to that of a single monolayer (*Supplementary Information Part VI*). This observation strongly suggests that the luminescence originates solely from the Au-Si interface and does not scale with the overall gold content beyond the initial monolayer. The low scattering activity and overdamped plasmonic properties of the 1.2 nm Au particles further rule out surface plasmon enhancement effects as the origin of the observed bright, broadband luminescence. Based on these findings, we conclude that gold PL cannot account for the bright visible emission observed with 1.2 nm Au particles on silicon, indicating the presence of a distinct mechanism.

The unique optical properties of the Si-particles interfacial region are further demonstrated in Figure 3. Figure 3a compares the Raman optical phonon signature from pure silicon and silicon with 1.2 nm, 15 nm Au and 1.2 nm Cu particles. When the 15 nm Au particles form larger clusters, creating accidental plasmonic hotspots, an enhancement of the Raman signal is observed. However, when the particles are assembled into intact monolayers, the Raman signal remains unchanged compared to bare silicon. In contrast, the presence of 1.2 nm Au or Cu particles strongly suppresses the inherent Raman signal from silicon. The suppression of the Raman signal suggests enhanced light absorption at the interface between bulk Si and particle, leading to reduced illumination of the underlying silicon volume.

Enhanced light absorption in silicon decorated with nanometer-sized particles has previously been reported and attributed to highly confined optical states.[24-26] According to Heisenberg's uncertainty principle, strong spatial confinement of light leads to a broadened distribution of photonic momenta, which, for sufficiently small particles, can span the entire Brillouin zone. [27-29] These momentum-broadened optical states enable photon-induced indirect transitions in silicon without phonon assistance, significantly accelerating the absorption process.

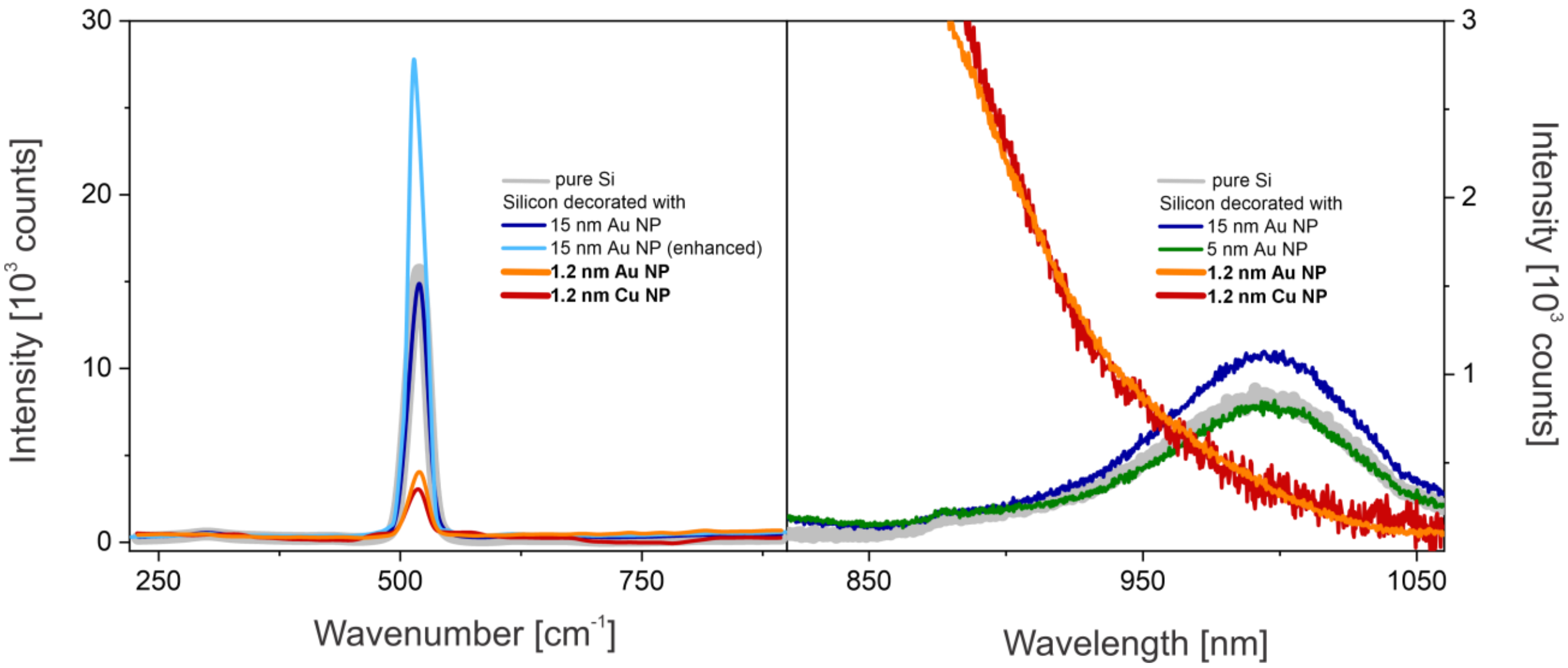

**Figure 3.** (a) The 521 cm$^{-1}$ Raman line of crystalline silicon. Larger particles show a modest surface enhancement, while areas with intact 1.2 nm particles show strong suppression of silicon’s Raman signal. Excitation using 532 nm, 0.5 mW, 0.75 NA. (b) PL in pure Si originates from phonon-assisted transitions from the bottom of the conduction band. The PL is fully suppressed when 1.2 nm particles are present.

From the observed suppression of the Raman signal, we find that the effective absorption coefficient near the Si surface has increased *up to three orders of magnitude*, consistent with experiments reported in a previous study.[26] This relaxed momentum conservation requirement is also evident in surface-enhanced Raman scattering (SERS) experiments on gold plasmonic antennas, where it contributes to the observed broad electronic Raman background of the metal [30-34]. We propose that these momentum-broadened optical states also play a crucial role in enhancing radiative rates for PL in silicon, leading to the observed bright emission across the visible spectrum.

Figure 3(b) shows that phonon-assisted radiative recombination from the bottom of the conduction band remains intact or even enhanced for 5 nm and 15 nm Au particles. However, when either Au or Cu single layer 1.2 nm particles are deposited onto the wafer surface, this recombination pathway is almost completely suppressed. This observation suggests that a new, highly efficient radiative channel emerges, one that outcompetes both intrinsic radiative and non-radiative relaxation processes near the band edge.

This new radiative channel is attributed to momentum-broadened optical states that enable direct recombination of conduction electrons with holes in the valence band along a broad range of momentum pathways. Figure 4a illustrates a representative example of the transition scenario involving holes near the Γ point only. Upon visible light excitation, electrons undergo ultrafast relaxation (<0.5 ps) to the bottom of the conduction band through electron–electron and electron–phonon scattering processes [4-6]. Full thermalization with the lattice is completed over longer timescales, typically few picoseconds and up to 1 nanoseconds at high density due to the formation of a hot-phonon bottleneck. Once thermalized, the photoexcited electron population in bulk silicon must remain trapped, depleting slowly via phonon-assisted radiative and intrinsic non-radiative relaxation processes on timescales of tens of microseconds.

This paradigm shifts entirely when confined photonic states are introduced. For a given spatial confinement $\sigma_r$, the photon acquires an expanded momentum spectrum $\sigma_k$ (Figure 4a, purple Gaussian profile) [26]. This relationship follows directly from the quantum uncertainty principle $\sigma_r \sigma_k = \left(n + 1/2\right)$, where n is the expectation value of the occupation number operator $\widehat{N} = a^\dagger a$ (see Supplementary Information Part VII for additional discussion). The expansion of photon momentum under sub-2 nm confinement leads to a dramatic enhancement in silicon's optical absorption, with estimated increases of up to three orders of magnitude [26]. In our current experiments, this effect is evidenced by the pronounced depletion of the intrinsic Si–Si Raman line observed beneath both Au and Cu nanoparticles with diameters of $d$ = 1.2 nm ($\sigma_r \sim d/2.355$=0.5 nm, $\sigma_k$=3 nm$^{-1}$). This observation has significant implications. Although the incident photon fluxes are modest, typically yielding an estimated conduction-band electron density of ~$10^{19}$ cm$^{-3}$ for freely propagating light, the enhanced absorption, driven by momentum-expanded photonic states ($\sigma_k$), is expected to proportionally increase the carrier density $n_{CB}$.

At sufficiently high electron densities in the conduction band (>$10^{20}$ cm$^{-3}$), the system enters a highly degenerate electron–hole plasma regime [35-37], characterized by a Fermi energy that rises well above the conduction band edge. In this regime, electrons begin to significantly fill the conduction band valleys, forcing the population to spread into higher-energy states. Representative electron distributions for $\sigma_k$ = 2 and 3 nm$^{-1}$ are shown in Figures 4b and 4c (see also Figure SF18).

Simultaneously, momentum-expanded photonic states enable a new radiative channel, with the transition probability $P(\sigma_k)$ defined as the projection of the Gaussian photon momentum distribution onto the conduction band dispersion profile. This probability is highest for electrons occupying elevated conduction band states (Figure 4b) and broadens significantly toward the band edge for photon confinement below 2 nm ($\sigma_k > 1.7\ nm^{-1}$; Figures 4c and SF19). The broadening of both the electron distribution and the radiative transition probability as a function of $\sigma_k$ results in a substantial spectral overlap between the two. This overlap establishes a viable pathway for radiative depletion of the conduction band and facilitates the emission of broadband radiation. Supporting this strongly confinement-dependent mechanism, we observe that the center of mass of the emission spectrum shifts markedly toward higher energies with increasing input flux (Figures SF20 and SF21). While the red side of the emission spectrum remains nearly unchanged with varying excitation power, the observed blue shift of 0.2–0.3 eV is consistent with expectations for electron densities exceeding $10^{21}\ cm^{-3}$ (Figure SF22; see discussions in Supplementary Information Part VIII). As expected, when the small particles within the focal spot melt or coalesce into larger clusters under intense and prolonged illumination, they can no longer sustain the required momentum expansion for these transitions. As a result, the system reverts to displaying the original Raman and phonon-assisted photoluminescence features characteristic of bulk silicon (Figures SF23 and SF24, Supplementary Information Part IX).

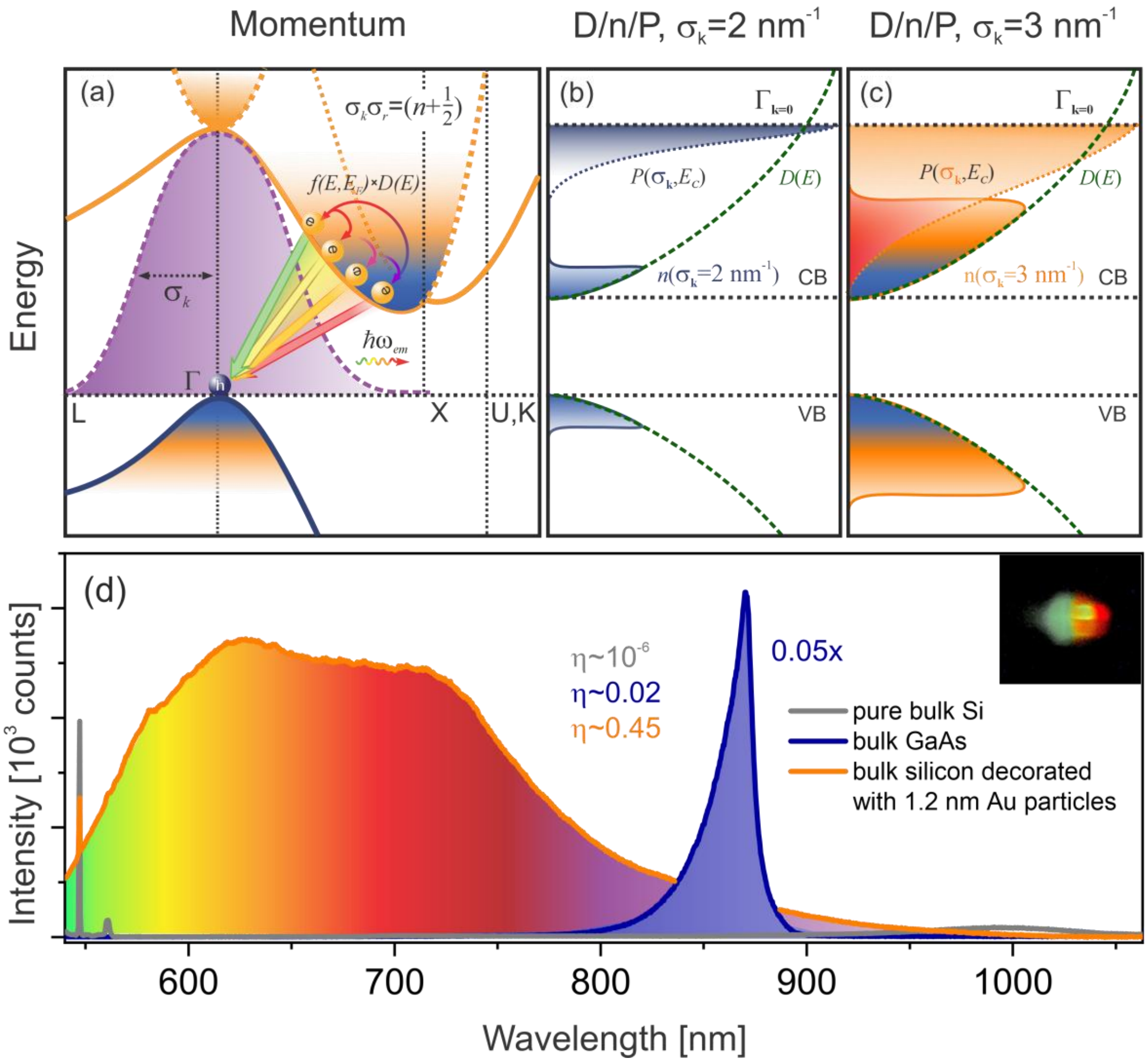

**Figure 4.** (a) Schematic illustration of optical transitions in bulk silicon decorated with sub-2 nm Au or Cu nanoparticles. Momentum-enhanced optical absorption drives a high-density electron gas into the conduction band. As the band fills, electron population redistributes toward higher-energy states. (b) Calculated electron distribution and corresponding transition probability function for $\sigma_k$ = 2 nm$^{-1}$. (c) Both functions broaden significantly with decreasing confinement size (i.e., increasing $\sigma_k$). For $\sigma_k > 2$ nm$^{-1}$, the resulting spectral overlap provides an efficient pathway for ultrafast radiative depletion of the conduction band population, giving rise to broadband emission. (d) Emission spectra of bare silicon, silicon decorated with 1.2 nm Au nanoparticles, and bulk GaAs measured under identical conditions. Comparison with GaAs luminescence indicates an external quantum efficiency exceeding 0.4 for the decorated silicon. *Inset*: Strong broadband emission from the decorated silicon wafer, visible to the naked eye. The emission is viewed through an Amici prism and recorded using an iPhone camera under 532 nm excitation with 10 s integration time (see Figure SF25, Supplementary Information Part XI).

Our results confirm that the newly enabled radiative transition pathways efficiently deplete the conduction band electron population in bulk silicon, effectively bypassing conventional radiative and non-radiative loss mechanisms and yielding high quantum efficiency emission. To quantify this efficiency, we calibrated the enhanced silicon emission intensity against the 865 nm photoluminescence peak of undoped bulk GaAs, used here as a reference standard (Figure 4e). Based on this analysis, we estimate an external quantum efficiency of approximately $\eta \approx 0.5$, comparable to that of direct bandgap GaAs ($\eta_{GaAs}$~0.72) [38,39] (see Table 1, Supplementary Information Part X for details). Despite being confined to the Si interface, this high quantum efficiency results in a sufficiently bright emission that is readily visible to the naked eye, as shown in Figure 4f.

These findings demonstrate that by carefully engineering confined optical states, silicon - traditionally a material with inherently low emission efficiency - can be transformed into a highly efficient light emitter. In this study, sub-2 nm Au or Cu nanoparticles were used to induce extreme photon confinement at the silicon surface, revealing that particle size, rather than chemical identity, is the key parameter governing this transformation. Importantly, the enhancement of light emission occurs without altering bulk silicon wafer, its crystal or electronic structures, indicating the observed emission enhancement arises primarily from the altered photonic environment. This simple and material-friendly approach may pave the way for the development of practical silicon-based light sources compatible with existing optoelectronic circuit fabrication processes. The concept of momentum-broadened optical states challenges the long-held assumption that silicon's optical properties are fundamentally constrained by its indirect bandgap. The results reported here open exciting new directions for the design and realization of advanced silicon-based optical components.

## Methods

*Synthesis of nanoparticles*

*Synthesis of 1.2 nm nanoparticles.* 300 µl of 100 mM aqueous nicotinamide adenine dinucleotide (NAD) solution was incubated with 80 mM $AuHCl_4$ at ambient temperature for 15 min. The mixture was added to 30 mL of freshly prepared aqueous solution containing 8 mM KOH and 1.5 mM $NaBH_4$ under constant high-speed stirring. An intensive dark brown color appeared, indicating nanoparticle formation. The solution was stirred for an additional 5-10 min. The particles were then centrifuged for 5 min at 20°C in two 15 mL 50 kDa Amicon-Ultra Centrifugal Filter Units at 4000 rpm. The filtrate was transferred into two 15 mL 10 kDa Amicon-Ultra Centrifugal Filter Units and centrifuged at 4000 rpm for 15 min. The retentate fractions (that did not pass through an ultrafiltration unit), containing concentrated particles (0.2–0.3 mL from each filtration unit), were pooled together. The pooled retentate was diluted into 15 mL of double-distilled water (DDW) and centrifuged in a 15 mL 10 kDa Amicon Ultra Centrifugal Filter Unit as described above. The centrifugation/dilution cycle was repeated four times to ensure complete removal of unbound NAD. After the last centrifugation/dilution cycle, the final retentate fraction (~0.2-0.3 mL) was collected. The absorbance of the particles at 420 nm was commonly equal ≈ 100 AU. The particles are stable and can be stored either at 4 ºC of 25 ºC for at least one month.

*Synthesis of 5 nm and 15 nm Au nanoparticles.* 5 nm and 15 nm Au nanoparticles (Au-NP's) were synthesized essentially as described in [40,41]. 30 mL of the 15-nm particles was centrifuged for 5 min at 20 °C in a 15 mL 100 kDa Amicon-Ultra Centrifugal Filter Units at 2000 rpm. The final retentate fraction (~0.2 mL) was collected. 30 mL of 5-nm Au-NP's was centrifuged for 15 min at 20 °C in a 15 mL 10 kDa Amicon-Ultra Centrifugal Filter Units at 4000 rpm. The final retentate fraction (~0.2 mL) was collected. The absorbance of both types of the particles at 520 nm was ~ 70 AU. The particles can be stored under ambient conditions for months.

*Deposition of nanoparticles on various surfaces*

*Deposition of 1.2 nm NAD-NP's on silicon.* A crystalline silicon wafer (<100>, undoped, 280 µm thickness) was treated with 5% HF for 5 min. The acid was thoroughly removed by rinsing with DDW, and the surface was dried by a flow of nitrogen gas. A 10-20 µL drop of NAD-NPs (OD ~

100 AU at 420 nm) in 0.3M KCL was applied to the surface and left on it for 15 hours in a humid atmosphere. The surface was then rinsed with cold DDW and dried with a nitrogen gas flow.

*Deposition of 1.2 nm nanoparticles on mica.* A 10-20 µL drop of NAD-NP's (OD ~ 100 AU at 420 nm) in 0.2M KCL was applied to a freshly cleaved mica and left on it for 15 hours in a humid atmosphere. The surface was then rinsed with cold DDW and dried with a nitrogen gas flow.

*Deposition of 15 nm and 5 nm Au nanoparticles on silicon.* A crystalline silicon wafer (<100>, undoped, 280 µm thickness) was treated with 5% HF for 5 min. The acid was thoroughly removed by rinsing with DDW, and the surface was dried by a flow of nitrogen gas. A 10-20 µL drop of 15-nm Au-NP's (OD ~ 50 AU at 520 nm) in 50 mM KCL and 5 mM Bis(p-sulfonatophenyl)phenylphosphine dihadrate dipotassium salt (BSPP) or 5-nm Au-NP's (OD ~ 50 AU at 520 nm) in 150 mM KCL and 5 mM BSPP was applied to the surface and left on it for 15 hours in a humid atmosphere. The surface was then rinsed with cold DDW and dried with a nitrogen gas flow.

*Reflection spectroscopy*

Reflection spectra were measured on a Cary-7000 universal measurement spectrometer (Agilent, US). The spectral bandwidth used for all measurements is 1 nm.

*Raman and PL micro-spectroscopy*

All micro-spectroscopy experiments were performed on a custom-modified microscopy system based on an InVia Renishaw system using 532 nm and 785 nm laser sources. The samples were illuminated with a 0.75 NA air objective (Leica), and emission was collected in the epi-configuration. The spectra were measured in expanded mode with a 1200 gr/mm diffraction grating used for spectral scanning with 0.6 $cm^{-1}$ resolution over the entire spectral region.

*Atomic Force Microscopy mapping*

Atomic Force Microscopy mapping was performed on a Solver PRO AFM system (NT-MDT Ltd.) in a semi-contact (tapping) mode using High Accuracy Non-Contact AFM probes from the PHA-NC series (ScanSens, Berlin, Germany). The images were "flattened" (each line of the image was fitted to a second-order polynomial, and the polynomial was then subtracted from the image line) with Nova image processing software (NT-MDT Ltd.). The images were analyzed using the following imaging software programs: WSxM Nanotec Electronica S.L (WSxM v4.0 Beta 10.0, Nanotec Electronica Ltd., Madrid, Spain) and SPIP software (MountainsSPIP®8, Image Metrology A/S, Hørsholm Denmark).

*HR-TEM Analysis of NAD-NPs*

HR-TEM images were acquired using a Thermo Fisher Scientific Talos F200i transmission electron microscope. Sample preparation involved depositing 4 µL of NAD-NPs solution (absorbance at 420 nm = 3 AU) onto an ultrathin (3–4 nm) carbon-coated copper grid. After 1 minute of incubation, excess solution was carefully removed by touching the edge of the grid with filter paper.

*Batch fluorescence spectroscopy*

Fluorescence spectroscopy of batch solutions containing compounds used in synthesis and deposition was performed using a Shimadzu RF6000 fluorescence spectrometer.

**Data availability**

The data is available from the corresponding author on reasonable request.

**Competing interest**

The authors declare no competing interests.

**Acknowledgements**

The authors thank Prof. Sasha Chernyshev, Prof. Paul H.M. van Loosdrecht, Prof. Ilya Krivorotov, Prof. Maxx Arguilla for fruitful discussions. D.A.F. and E.O.P acknowledge funding from Chan Zuckerberg Initiative 2023-321174 (5022) GB-1585590, NSF 2025-2434622, and SBIR grant through Space Agency c029-102623. C.G. acknowledges funding from the Swiss National Science Foundation (SNSF) under grant 214993. D.A.F. thanks Yulia Davydova for support and help during this study. D.A.F. dedicates this work to beloved friend Fiona.